\documentclass[conference,10pt]{IEEEtran}
\IEEEoverridecommandlockouts

\usepackage{cite}
\usepackage{subcaption}
\usepackage{graphicx}
\usepackage{dcolumn}   % needed for some tables
\usepackage{bm}        % for math
\usepackage{bbm}
\usepackage[colorlinks=true, 
urlcolor=blue,citecolor=blue,anchorcolor=blue,breaklinks=true]{hyperref}
\usepackage{dsfont}
\usepackage{mathtools}
\usepackage{amsthm, thmtools}
\usepackage[normalem]{ulem}
\usepackage{amsmath}
\usepackage{empheq}
\usepackage[ruled]{algorithm2e}
\usepackage{xcolor}
\usepackage[normalem]{ulem} %temp for \sout
\usepackage{comment}
\usepackage[capitalise]{cleveref}

\newcommand{\ket}[1]{\left|#1\right\rangle}
\newcommand{\bra}[1]{\left\langle#1\right|}

\newcommand{\avg}[1]{\left\langle#1\right\rangle}
\usepackage{mathtools}

\usepackage{bm}
\let\vec\bm 

\begin{document}
\title{QAOA Parameter Transfer for Hypergraphs
\thanks{This material is based upon work supported by the U.S. Department of Energy, Office of Science, Office of Advanced Scientific Computing Research under Award Number 89243024SSC000129 and under field work proposal ERKJ445.  This research used resources of the Compute and Data Environment for Science (CADES) at the Oak Ridge National Laboratory, which is supported by the Office of Science of the U.S. Department of Energy under Contract No. DE-AC05-00OR22725. This manuscript has been authored by UT-Battelle, LLC, under Contract No. DE-AC0500OR22725 with the U.S. Department of Energy. The United States Government retains and the publisher, by accepting the article for publication, acknowledges that the United States Government retains a non-exclusive, paid-up, irrevocable, world-wide license to publish or reproduce the published form of this manuscript, or allow others to do so, for the United States Government purposes. The Department of Energy will provide public access to these results of federally sponsored research in accordance with the DOE Public Access Plan.}
}

\author{
\IEEEauthorblockN{Lucas T. Braydwood}
\IEEEauthorblockA{\textit{Quantum Artificial Intelligence Laboratory} \\
\textit{NASA Ames Research Center}\\
Moffett Field, CA USA \\
lucas.t.brady@nasa.gov}
\and
\IEEEauthorblockN{Phillip C. Lotshaw}
\IEEEauthorblockA{\textit{Quantum Information Science Section} \\
\textit{Oak Ridge National Laboratory}\\
Oak Ridge, TN USA\\
lotshawpc@ornl.gov}
}

\maketitle

\begin{abstract}
    Variational Quantum Algorithms, including the Quantum Approximate Optimization Algorithm (QAOA), have shown promise in solving optimization problems but rely on costly variational loops that can themselves be hard optimization problems.  Many methods have been proposed to mitigate this variational cost, with one of the most common being parameter transfer and concentration where variational parameters for one problem instance or for an average over problem instances can be used as a good set of parameters for another instance.  Methods exist for reweighting these parameters based off graph degree and edge weights, but there has been little work on how to do this reweighting to handle higher locality problems where the graph structure turns into a hypergraph structure.  In this paper, we analytically derive parameter reweighting rules to transfer parameters between different locality hypergraphs, resulting in a reweighting for the mixing terms in the Hamiltonian which have previously not been considered.  These analytics rely on three cycle-free and low-circuit-depth assumptions, but numerics indicate that the results can be used even when these assumptions are not satisfied. The numerics obtain high quality results across a diverse set of hypergraphs with locality less than or equal to five, improving on previous relations that do not reweight the mixing terms.
\end{abstract}


\section{Introduction}

There is promise for quantum algorithms to aid with optimization tasks, whether through the use of adiabaticity \cite{farhi2000}, variational approaches \cite{Farhi2014,hadfield2019quantum,Peruzzo2014}, control theory \cite{Magann_2022,Brady_2024}, or other mechanisms \cite{Jordan_2025}.  While promising, many of these algorithms rely on either long runtimes or long variational stretches to reach good fidelity with low energy solutions to the optimization problem.  For instance, the Quantum Approximate Optimization Algorithm (QAOA) \cite{Farhi2014,hadfield2019quantum} can produce good overlap with the ground state, but in its full form, it requires a variational optimization that is NP-Hard in general.

Parameter Transfer or Parameter Concentration is a phenomenon in QAOA analysis where parameters from solving QAOA on one problem instance can be used to produce an optimized or close-to-optimized circuit for another problem instance \cite{lotshaw2021empirical ,lotshaw2023approximate,Wurtz_2021,zhou2020quantum}.  Parameter concentration has been shown analytically for the Sherrington-Kirkpatrick model in the infinite-$n$ limit \cite{Farhi_2022}, for smaller $n$ with drastically simplified toy problems \cite{Akshay_2021}, and for sufficiently low $p$ over structured sets of random graphs \cite{brandao2018fixed}.  For more realistic problem sizes and settings, there has been some work analytically and numerically to justify observed parameter concentration for quadratic optimization problems \cite{galda2023,Shaydulin_2023,xu2025}.  All of these techniques are employed and tested on quadratic problems, but there is little work on higher locality problems. A notable exception is Ref.~\cite{Sureshbabu_2024}, which proposed a formula for rescaling based on edge weights of hypergraphs, though to our knowledge this has not been tested numerically at locality three or greater.

This paper will focus on these parameter transfer techniques in the context of higher locality problems.  There are techniques for reweighting parameters to move to graphs with different densities and weights, but these techniques exclusively focus on reweighting the $\gamma$ parameters that govern the evolution of the cost Hamiltonian.  We find that for higher-locality problems, the reweighting in $\gamma$ is not enough and that the $\beta$ parameters that control the mixing Hamiltonian need to be altered as the locality changes.

The analytic techniques that have been involved in the past fall into one of three categories.  The first, which uses asymptotic properties of the system, is not easily derived in realistic settings for most problems.  The second techniques is much more relevant.  It relies on the fact that the objective function we are trying to optimize for in QAOA is a linear combination of correlators of edges.  The structure of the standard QAOA ansatz (with a transverse field mixer) means that for each layer in the ansatz, $p$, corresponds to additional graph locality that needs to be considered.  So at $p=1$ we need to consider nodes that share an edge with the nodes in the edge being considered, and at $p=2$ we need to consider edges one further out and so on.  This portion does generalize to hypergraphs and hyperedges in higher locality problems, but the possible hypergraph structures increase in number much faster than the possible graph features.  The last technique is to consider triangle-free graphs where analytics can be worked out and then generalizing from there.  This is the technique we will focus on, but it will require generalization of the notion of triangle-free to hypergraphs.

\section{QAOA and Setup}

The Quantum Approximate Optimization Algorithm (QAOA) \cite{Farhi2014,hadfield2019quantum} is a variational algorithm that applies an alternating bang-bang pattern of problem and mixer Hamiltonians to a simple starting state in order to generate a state that minimizes the energy with respect to the problem Hamiltonian.

The structure of QAOA is to create an ansatz state of the form
\begin{equation}
    \ket{\psi(\vec{\beta},\vec{\gamma})} = e^{-i\beta_p\hat{B}}e^{-i\gamma_p \hat{C}}
    \dots
    e^{-i\beta_1\hat{B}}e^{-i\gamma_1 \hat{C}}
    \ket{\varphi_0},
\end{equation}
where the mixer is often taken to be
\begin{equation}
    \hat{B} = -\sum_{i=1}^n \sigma_i^{(x)}
\end{equation}
with the initial state $\ket{\varphi_0}$ is its ground state with all the $n$ qubits in the $\ket{+}$ state.  We will use these normal conventions, but more advanced mixers are possible and can lead to favorable properties \cite{hadfield2019quantum,Wang_2020}.

There are $p$ layers here with variational parameters $\vec{\gamma}$ and $\vec{\beta}$.  The problem Hamiltonian, $\hat{C}$, encodes the problem we are trying to minimize into its eigenspectrum.  The typical structure would be to encode a classical optimization function along the diagonal of the Hamiltonian using products of Pauli-Z operators.

In this paper, we will assume that the problem Hamiltonian takes on a form
\begin{equation}
    \label{eq:prob_ham_klocal}
    \hat{H}^{(T)} = \sum_{\alpha=1}^m w_\alpha \bigotimes_{j\in Q_\alpha} \sigma_j^{(z)},
\end{equation}

where $\alpha$ indexes over terms in the Hamiltonian.  From a graph point of view, this can be interpreted as solving Max-$k$-XORSAT on the hypergraph determined by $n$ nodes connected by hyperedges defined by $Q_\alpha$ which are the sets of nodes in the hyperedge $\alpha$.  This is the natural way of encoding binary optimization problems into Hamiltonians, and any unconstrained binary optimization problem can be encoded in this form (constraints can be handled in this form using penalties).

In this paper, we start with this hypergraph problem and work to derive concentration and transfer results for the $\vec{\beta}$ parameters that will take them to sectors with different or mixed hyperedge localities, $k_\alpha = |Q_\alpha|$.

\section{Analytics Setup}

First we want to write down the general forms for the expectation values of strings of Pauli-$Z$ matrices under QAOA for $k$-local Hamiltonians.  A general form of this is too complicated to be helpful, so we need to narrow the scope.  We will work with $p=1$ QAOA to make the analytic expressions tractable.  Once we have these, we can construct the form of the entire energy function by summing over all terms in the Hamiltonian.  

A recent paper \cite{Brady2025} provides an expression for a general string of Pauli-Zs in the QAOA cost function, assuming a simple transverse field mixer.  Consider that we have a string $Q_{\alpha}$ that contains nodes that we want the expectation value of.  We will use $N(\alpha)$ to denote the nodes that share a hyperedge with any bits in this set, and $C_\alpha(z)$ will refer to the parts of the cost function that contain any of the nodes from $Q_\alpha$.  With this, the most general form we can write down is
\begin{align}
    \label{eq:Jgeneral}
    J_{Q_\alpha}(\beta,\gamma)&=\bra{\psi(\vec{\beta},\vec{\gamma})}\left[\prod_{j\in Q_\alpha}\sigma_j^{(z)}\right]\ket{\psi(\vec{\beta},\vec{\gamma})}\\\nonumber &= \frac{1}{2^{d_\alpha+2*k_\alpha}}
    \sum_{\substack{z_k=\pm 1\\k\in N(\alpha)}}
    \sum_{\substack{z_j=\pm 1\\j\in Q_\alpha}}
    \sum_{\substack{z'_j=\pm z_j\\j\in Q_\alpha}}
    e^{i\gamma (C_\alpha(z')-C_\alpha(z))}\\\nonumber
    &
    \times\prod_{j\in Q_\alpha}\left(e^{2i\beta}(-1)^{\frac{1-z_j}{2}} + 
            e^{-2i\beta}(-1)^{\frac{1-z'_j}{2}}\right).
\end{align}
Similarly, $d_\alpha = |N(\alpha)|$ and $k_\alpha = |Q_\alpha|$.  Note that this expression corrects a sign error found in the original version of Ref.~\cite{Brady2025}.  We will use this as a starting point for our analytics.

\begin{figure}
\includegraphics[height=6cm,width=\columnwidth,keepaspectratio]{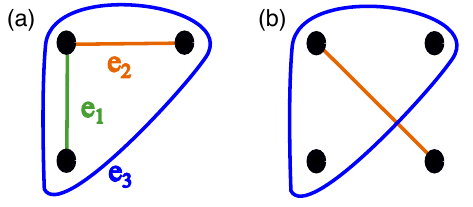}
\caption{Berge cycles generalize the idea of a triangle to hypergraphs.  (a) An example hypergraph for which each vertex is connected back to itself along a sequence of edges $e_1,e_2$ and the hyperedge $e_3$, corresponding to Berge cycles of length three.  The hypergraph also contains Berge cycles of length two.  (b) Example hypergraph that does not contain any Berge cycle.
}
\label{Berge cycle}
\end{figure}

\section{Acyclic Hypergraphs}

A technique that is often employed in parameter transfer results on graphs is to consider graphs that are triangle free which is a limit where the analytics can be worked out \cite{Shaydulin_2023}.  If we want to generalize this to hypergraphs, we could look at the notion of Berge cyclicity and require that our hypergraph have no Berge cycles of length less than four.  If we were to consider higher $p$, we would have to extend the restricted length to continue using the benefits of this condition. Example graphs with and without Berge cycles are shown in Fig.~\ref{Berge cycle}.

This allows us to make a number of simplifications in the equations.  Let us start by revisiting Eq.~(\ref{eq:Jgeneral}).  The important thing to note now is that the neighborhood $N(\alpha)$ now breaks up cleanly into a series of neighborhoods around each vertex in $Q_{\alpha}$, $N'(j)$, where the prime indicates that this neighborhood does not include any of the other vertices in $Q_{\alpha}$.  These sub-neighborhoods do not have any overlap, and the relevant sums will now break down into very clean portions.  First we will show this breaking down into neighborhoods, using $C_j'(z)$ to denote those elements of $C(z)$ that contain the node $j$ with the prime excluding the hyperedge we are trying to measure, $Q_\alpha$.

\begin{align}
    J_{Q_\alpha}(\beta,\gamma) &= \frac{1}{2^{d_\alpha+2*k_\alpha}}
    \sum_{\substack{z_j=\pm 1\\j\in Q_\alpha}}
    \sum_{\substack{z'_j=\pm z_j\\j\in Q_\alpha}}
    e^{i\gamma (Q_\alpha(z')-Q_\alpha(z))}\\\nonumber
    &
    \times \prod_{j \in Q_\alpha}\left[\sum_{\substack{z_k=\pm 1 \\k \in N'(j)}}e^{i\gamma (C'_j(z')-C_j(z))}\right]\\\nonumber
    &
    \times\prod_{j\in Q_\alpha}\left(e^{2i\beta}(-1)^{\frac{1-z_j}{2}} + 
            e^{-2i\beta}(-1)^{\frac{1-z'_j}{2}}\right).
\end{align}
Here $Q_\alpha(z) = w_\alpha\prod_{j\in Q_{\alpha}} z_j$, where $w_\alpha$ is the weight of the hyperedge $\alpha$.

Let's take a close look at $C'_j(z)$ in this acyclic environment.  It will be a sum of terms $r \in C'_j(z)$.  Each of these $r$ will have a weight $w_r$ as well as $z_j$ and a collection of other bits that do not appear anywhere else in the above expression.  Therefore, we are looking at a general set of sums that look like
\begin{equation}
    \left[\prod_{i=1}^{k_r-1}\sum_{z_i=\pm1}\right]e^{i\gamma w_r (z'_j-z_j) \prod_{i=1}^{k_r-1}z_i}.
\end{equation}
There is no reason why we cannot complete these sums now since these variables appear nowhere else.  Since the exponential contains a product of plus or minus ones with even distribution, we can see that this comes out to be 
\begin{equation}
    2^{k_r-1}\cos\left(\gamma w_r (z'_j-z_j) \right).
\end{equation}

All of the $2^{k_r-1}$ from all of the terms will exactly cancel out the factors of 2 from the neighboring bits in the original expressions, $2^{d_\alpha}$.  So if $E'(j)$ are the set of hyperedges that contain node $j$, excluding the $\alpha$ hyperedge, then we can simplify this down to
\begin{align}
    J_{Q_\alpha}(\beta,\gamma) &= \frac{1}{2^{2*k_\alpha}}
    \sum_{\substack{z_j=\pm 1\\j\in Q_\alpha}}
    \sum_{\substack{z'_j=\pm z_j\\j\in Q_\alpha}}
    e^{i\gamma (Q_\alpha(z')-Q_\alpha(z))}\\\nonumber
    &
    \times \prod_{j \in Q_\alpha}\prod_{r\in E'(j)}\cos\left(\gamma w_r (z'_j-z_j) \right)\\\nonumber
    &
    \times\prod_{j\in Q_\alpha}\left(e^{2i\beta}(-1)^{\frac{1-z_j}{2}} + 
            e^{-2i\beta}(-1)^{\frac{1-z'_j}{2}}\right).
\end{align}

For the next step, we evaluate the rightmost sum in the top line by considering all possible subsets of the nodes in our hyperedge of interest, $q\subseteq$ $Q_{\alpha}$.  For each subset $q$, we evaluate those nodes in the subset such that $z'_j=-z_j$, and those elements of $Q_{\alpha}$ that were not in $q$ evaluate with $z'_j = z_j$.  Then instead of a sum over bit evaluations, we are left with a sum over these subsets. 
\begin{align}
    J_{Q_\alpha}(\beta,\gamma) &= \frac{1}{2^{k_\alpha}}
    \sum_{\substack{z_j=\pm 1\\j\in Q_\alpha}}
    \sum_{q\subseteq Q_\alpha}
    e^{-i\gamma Q_\alpha(z)(1-(-1)^{k_q})}\\\nonumber
    &
    \times i^{k_q}\cos^{k_{\alpha}-k_q}(2\beta)\sin^{k_q}(2\beta)\\\nonumber
    &
    \times \prod_{j \in q}\prod_{r\in E'(j)}\cos\left(2 \gamma w_r \right) \\\nonumber
    &
    \times\prod_{j\in Q_\alpha}(-1)^{\frac{1-z_j}{2}}.
\end{align}
In general, we are using $k_a = |a|$ to mean the size of the set $a$ with $k_{\alpha} = |Q_\alpha|$ and $k_{q} = |q|$. 

Now we are going to evaluate the first sum in the expression in a similar way by considering subsets $p \subseteq$ $Q_\alpha$ where for each node, $j$, in the subset, we set $z_j = -1$, and for each node, $i$, not in a given subset, set $z_i = +1$:
\begin{align}
    J_{Q_\alpha}(\beta,\gamma) &= \frac{1}{2^{k_\alpha}}
    \sum_{p\subseteq Q_\alpha}(-1)^{k_p}
    \sum_{q\subseteq Q_\alpha}i^{k_q}\cos^{k_{\alpha}-k_q}(2\beta)\sin^{k_q}(2\beta)
    \\\nonumber
    &
    \times e^{-i\gamma w_{\alpha} (-1)^{k_p}(1-(-1)^{k_q})}\\\nonumber
    &
    \times \prod_{j \in q}\prod_{r\in E'(j)}\cos\left(2 \gamma w_r \right).
\end{align}
 The only thing we care about here from the $p$ subsets is how large they are, so we can replace this with a binomial distribution easily
\begin{align}
    J_{Q_\alpha}(\beta,\gamma) &= \frac{1}{2^{k_\alpha}}
    \sum_{q\subseteq Q_\alpha}i^{k_q}\cos^{k_{\alpha}-k_q}(2\beta)\sin^{k_q}(2\beta)
    \\\nonumber
    &
    \times \prod_{j \in q}\prod_{r\in E'(j)}\cos\left(2 \gamma w_r \right)\\\nonumber
    &
    \times \sum_{l=0}^{k_\alpha}{k_\alpha \choose l}(-1)^{l}e^{-i\gamma w_{\alpha} (-1)^{l}(1-(-1)^{k_q})}.
\end{align}

And then this binomial sum is straightforward to do.  Assuming that we do not have $k_\alpha = 0$ (which we do not since $\alpha$ represents a non-trivial hyperedge), we are left with
\begin{align}
    J_{Q_\alpha}(\beta,\gamma) &=  
    -\sum_{q\subseteq Q_\alpha}i^{k_q+1}\cos^{k_{\alpha}-k_q}(2\beta)\sin^{k_q}(2\beta)\\\nonumber
    &
    \times \sin(\gamma w_{\alpha}(1-(-1)^{k_q}))
    \\\nonumber
    &
    \times \prod_{j \in q}\prod_{r\in E'(j)}\cos\left(2 \gamma w_r \right).
\end{align}

Note importantly that if $k_q$ is even, this is imaginary, but the sine on the second line also contributes zero.  So only the real elements with odd $k_q$ will end up contributing.

\begin{align}
    J_{Q_\alpha}(\beta,\gamma) &= -\sin(2\gamma w_{\alpha})\\\nonumber
    &
    \times
    \sum_{\substack{q\subseteq Q_\alpha\\k_q~\text{odd}}}(-1)^{(k_q+1)/2}\cos^{k_{\alpha}-k_q}(2\beta)\sin^{k_q}(2\beta)\\\nonumber
    &
    \times \prod_{j \in q}\prod_{r\in E'(j)}\cos\left(2 \gamma w_r \right).
\end{align}

This result matches simulations of three-cycle free hypergraphs under QAOA.

\subsection{$\beta$ Fixing}

If we are going to want any hope of finding an optimal $\beta$ that is independent of $\gamma$, we are going to have to make an assumption that these angles are small.  To start with, we will assume $\gamma$ is a small quantity and expand this to second order in $\gamma$ so that 

\begin{align}
    J_{Q_\alpha}(\beta,\gamma) &= -2\gamma w_{\alpha}\\\nonumber
    &
    \times
    \sum_{\substack{q\subseteq Q_\alpha\\k_q~\text{odd}}}(-1)^{(k_q+1)/2}\cos^{k_{\alpha}-k_q}(2\beta)\sin^{k_q}(2\beta)
    \\\nonumber&
    +\mathcal{O}(\gamma^3).
\end{align}

Since we no longer have any dependence on $q$ other than its size, this simplifies dramatically.  We can employ multi angle formulas for trigonometric functions to get
\begin{align}
    J_{Q_\alpha}(\beta,\gamma) &= -2\gamma w_{\alpha} 
    \sin(2 k_\alpha \beta)+\mathcal{O}(\gamma^3).
\end{align}

The quantity that we actually want to optimize is
\begin{align}
    \avg{\hat{C}} &= \sum_{\alpha} w_\alpha J_{Q_\alpha}\\\nonumber
    &  = -2\gamma \sum_{\alpha} w_{\alpha}^2 \sin(2 k_\alpha \beta)+\mathcal{O}(\gamma^3).
\end{align}

Just to sanity check with our numeric findings, if all terms have the same locality $k$, this does indeed have its first optimum at $\beta = \frac{\pi}{4k}$

We can do an expansion here where we expand each sine around its first peak in an attempt to find the location of the combined peaks
\begin{align}
    \avg{\hat{C}} \approx -2\gamma \sum_{\alpha} w_{\alpha}^2 (1-\frac{1}{2}(2k_\alpha \beta-\frac{\pi}{2})^2).
\end{align}

If we take derivatives and try to find the location of the first combined peak is
\begin{equation}
    \label{eq:beta_transfer}
    \beta_* \approx \frac{\pi}{4}\frac{\sum_\alpha w_\alpha^2  k_\alpha}{\sum_\alpha w_\alpha^2 k_\alpha^2}
\end{equation}

In many settings, the results from Eq.~({\ref{eq:beta_transfer}) can be used directly for the $p=1$ $\beta$ value.  To use this for transfer suppose we have a reference $\vec{\beta}$ value (or vector of beta values for $p>1$), $\vec{\beta}^{(r)}$ where the reference hypergraph problem had a value of Eq.~({\ref{eq:beta_transfer}) evaluated to $\beta_*^{(r)}$.  Then if we want to know what the transferred $\vec{\beta}$ values are for a hypergraph that has a value of Eq.~({\ref{eq:beta_transfer}) of $\beta_*$, then we would calculate:
\begin{equation}
    \vec{\beta} = \frac{\beta_*}{\beta^{(r)}_*}\vec{\beta}^{(r)} 
\end{equation}
\section{Numerics}

We will use the $\beta$ transfer rule describe above in Eq.~\ref{eq:beta_transfer}, reweighting angles from one graph to another using this factor.  For the $\gamma$ angles, we will use a reweighting scheme similar to that used elsewhere in the literature.  Our hypergraph instances all include weights equal to $\pm1$, so reweighting schemes do not need to account for edge weight variations.  Still, it is typical to reweight $\gamma$s based on the average degree of the graph \cite{Shaydulin_2023,Wurtz_2021,Sureshbabu_2024}.  For instance, if we have a reference $\gamma^{(r)}$, then our reweighted
\begin{equation}
    \label{eq:gamma_reweighting}
    \gamma = \frac{\gamma^{(r)}}{\sqrt{D}},~~~~~D = \frac{\sum_{k}k m_k}{n},
\end{equation}
where $n$ is the number of qubits and $m_k$ are the number of terms in the hypergraph with locality $k$.
Future work will be directed to refining the $\gamma$ reweighting as well.

Our hypergraph instances were generated using the \texttt{fast\_random\_hypergraph} routine in the CompleX Group Interactions library \texttt{xgi}.  The routine generates a hypergraph with distinct edge probabilities at each specified locality. When using the routine, we ensure the resulting graphs is connected by repeating until a connected graph is found.  We generated a random set of hypergraphs with $N=14$ vertices and with (locality two) edge probabilities $p_2 \in \{0.0,0.1,0.2,0.3\}$.  We independently assigned hyperedges of locality $k$ using probabilities $p_k \in \{0.0,0.1r_k,0.2r_k,0.3r_k\}$ that were rescaled by factors $r_k =  {N \choose 2}/{N \choose k}$ to yield numbers of expected hyperedges that are the same as the number of edges expected from the various $p_2$.   For $p_1$, we use only three choices $p_1 \in \{0.0,0.1r_1,1\}$ as the previous rescaling procedure yields probabilities exceeding one.  We independently assign each $p_k$ up to maximum locality $k=5$, removing the three trivial cases where $p_2=p_3=p_4=p_5=0$, yielding $3\times 4^4 -3 = 765$ combinations of the $p_k$.  For each combination of the $p_k$ we generate four random instances, yielding 3,060 graphs total in our dataset.

We generate results for QAOA using three different schemes.  The first is a full variational form of QAOA based on independent optimizations of random initial angles using the \texttt{NLOPT} implementation of the low-storage Broyden–Fletcher–Goldfarb–Shanno algorithm. For each graph, we performed 100 of these optimizations at $p=1$ and 1,000 at $2 \leq p \leq 5$ to determine the best variational angles for each $p$. We also used boot-strapping methods to go from low to high $p$ \cite{Zhou_2020,Pagano_2020} as well as confirming optimality of the $p=1$ results using a full grid search.  For each hypergraph, at each $p$ we take the results from the best method and treat that as the optimized form of QAOA.

For our reweighting results, we start from optimized QAOA angles, available in the literature \cite{Wurtz_2021}.  These angles were derived for three-regular graphs, which are a much more specialized case than we consider here, meaning that these angles are not optimized for our use cases.  We consider reweighting using either just $\gamma$ reweighting from Eq.~(\ref{eq:gamma_reweighting}) or reweighting using both $\gamma$ and our $\beta$ reweighting.  While these parameter reweighting factors were derived for $p=1$, we will assume similarity at higher $p$ and use these factors to reweight all angles between graphs at higher circuit depth.

\begin{figure}
\begin{center}
\includegraphics[width = 0.48\textwidth]{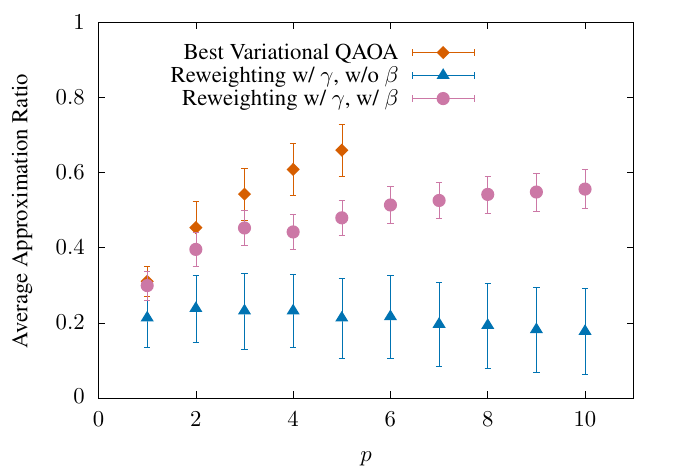}
\end{center}
\caption{The average approximation ratio across all our hypergraph instances as a function of QAOA circuit depth, $p$.  We plot the results for the best found QAOA angles, as well as reweighting of reference angles from the literature \cite{Wurtz_2021} using the $\gamma$ reweighting from Eq.~(\ref{eq:gamma_reweighting}) with and without the $\beta$ reweighting from Eq.~(\ref{eq:beta_transfer})
}
\label{fig:all_hypergraphs}
\end{figure}

\begin{figure}
\begin{center}
\includegraphics[width = 0.48\textwidth]{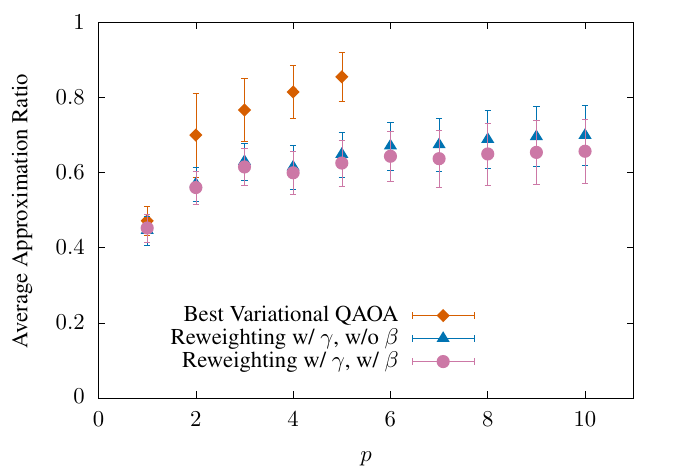}
\end{center}
\caption{The average approximation ratio across just the graph instances, including only terms with locality 2 or less, as a function of QAOA circuit depth, $p$.  We plot the results for the best found QAOA angles, as well as reweighting of reference angles from the literature \cite{Wurtz_2021} using the $\gamma$ reweighting from Eq.~(\ref{eq:gamma_reweighting}) with and without the $\beta$ reweighting from Eq.~(\ref{eq:beta_transfer})
}
\label{fig:1-2_graphs}
\end{figure}

In Figs.~\ref{fig:all_hypergraphs}~\&~\ref{fig:1-2_graphs}, we present our resulting numerics, looking at the average approximations ratios, defined as the achieved energy divided by the true ground state energy (as determined through exact diagonalization), as a function of QAOA circuit depth, $p$.  We plot the best found energy using a full variational QAOA as well as the energies achieved using our reweighting schemes (both with and without $\beta$ reweighting), starting from known reference angles \cite{Wurtz_2021} for 3-regular graphs.

Fig.~\ref{fig:all_hypergraphs} shows results for all our hypergraph instances where it is clear that the $\beta$ reweighting has a large improvement over not reweighting $\beta$s.  Just using $\gamma$ reweighting leads to an approximation ratio that decreases with circuit depth, indicating that we are far away from a minimum in the control landscape, but including our new $\beta$ reweighting leads to steady improvement with circuit depth, even past the circuit depths that were reasonable to simulate with a full variational circuit.  To contrast to this, we also include Fig.~\ref{fig:1-2_graphs} which only includes graphs with terms of locality 1 and 2.  Here, the $\beta$ reweighting has a minimal effect and is very slightly deleterious, likely due to interactions with the self-weighting terms in the Hamiltonian.  Our analytic derivation of the $\beta$ reweighting makes some assumptions that are tighter for higher locality and looser for lower locality.  Namely, we assumed in our formula that the peaks of the various oscillations were close together so that we could expand for small $\gamma$ deviations from each peak, and this assumption is more accurate for higher locality terms with similar localities.

\section{Conclusion}

Variational quantum algorithms require extensive time and resources to optimize their parameters, so any shortcut to this process can immediately lead to better algorithms.  In the case of QAOA, methods have been developed and tested favorably \cite{Shaydulin_2023,Wurtz_2021,Sureshbabu_2024,Farhi_2022,galda2023,xu2025} to transfer variational parameters from one graph structure to another.  These parameters need to be reweighted when the graph degree changes or when the graph weights change.  This paper outlines that reweighting must also be considered when the locality of the graph changes into a hypergraph.  Specifically, the $\beta$ angles, that govern the evolution of the mixer Hamiltonian, must be modified for hypergraphs which has not been considered before for graphs.

Our analytic derivation of the $\beta$ scaling works by assuming there are no Berge cycles of a certain length for $p=1$ QAOA and than analytically calculates optimal angles.  Numerically, this procedure works favorably but is still a ways off from the performance of a full variational QAOA.  Additional work is needed to identify other places where parameter transfer can be improved to ensure that a non-variational form of QAOA can be competitive. One possibility is that a $\gamma$ rescaling factor could be derived based on our results, to account for additional details of the hypergraph structure.

Additionally, further work is needed to identify conditions under which this kind of parameter transfer fails and what analytic conditions lead to it holding.  Relying on acyclic graphs for the analytics does not noticeably effect the performance of those analytics in regimes with cycles, as most of our numerics have.  Random hypergraphs tend to not have high overlap with computationally hard problems, so study specifically in hypergraphs corresponding to hard optimization problems may be fruitful for future work.

\section*{Acknowledgments}

The authors thank Stuart Hadfield, Titus Morris, Ryan Bennink, and Filip Maciejewski for discussions and feedback.
\bibliographystyle{IEEEtran}
\bibliography{refs}

\end{document}